\documentclass[prb,twocolumn,showpacs,preprintnumbers,amsmath,amssymb]{revtex4}

\usepackage{graphicx}

\begin{document}


\title{Substrate temperature changes during MBE growth of GaMnAs}

\author{V.~Nov{\'ak}}
\email{vit.novak@fzu.cz}
\author{K.~Olejn{\'{i}}k, M.~Cukr, L.~Smr\v{c}ka, Z.~Reme\v{s}}
\author{J.~Oswald}
\affiliation{%
Institute of Physics AS CR, Cukrovarnick\'a 10,
162~53 Praha 6, Czech Republic}

\date{\today}

\begin{abstract}
Remarkably big increase of the substrate temperature during
the low-temperature MBE growth of GaMnAs layers is observed
by means of band gap spectroscopy. It is explained and
simulated in terms
of changes in the absorption/emission characteristics 
of the growing layer. Options for the temperature variation
damping are discussed.
%
\end{abstract}


\maketitle

\section{Introduction}

The growth temperature is an important parameter determining 
the quality of the MBE grown ferromagnetic GaMnAs layers.
Its optimum value depends on the Mn doping level.\cite{Ohno,Sadowski,Camp03}
With increasing Mn density the growth temperature has to be reduced in order 
to prevent the Mn precipitation. At the same time, however, a certain
minimum temperature must not be underrun in order to maintain the 
2D growth. The growth temperature adjustment is critical especially
at high Mn doping levels where the temperature range of the 2D growth 
becomes narrow. Unfortunately, determining the substrate temperature
at this temperature region is biased by a considerable uncertainty:
the GaAs substrate is transparent for a low-temperature pyrometry, and
a thermocouple can only weakly be linked to a rotating substrate
holder. Moreover, it is difficult to eliminate a substantial delay 
in the temperature regulation. 

Application of the diffuse reflectance spectroscopy\cite{BESthermo}
(or band edge spectroscopy, BES) not only allows 
one to overcome the above problems, but also reveals remarkable
temperature changes connected with the massive Mn doping.
Similar observation has already been reported in connection
with growth of smaller band gap layers on larger band gap 
substrates,\cite{Shanabrook} and with heat flux transients
due to opening/closing the effusion cell shutters.\cite{Thompson}
The latter problem is certainly involved in our case, too. 
However, as we show further, yet another mechanism takes
effect, becoming essential when growing heavily doped
samples mounted in In-free sample holders.

\section{Experiment} 
\label{exper}

The growth experiments were performed in Veeco Gen II
MBE system. A semiinsulating $500\mu$m thick GaAs substrate 
with 2-inch diameter was mounted into a molybdenum 
In-free sample holder, fixed mechanically at its edge. 
No diffuser plate was used between the substrate and
the heater. Distance between the heater (2-inch PBN plate 
with tungsten filament inside) and the substrate was about 
10 mm. The band-edge spectrometer (kSA BandiT) was mounted on 
the central pyrometer port, normal to the 
substrate. In this arrangement the thermal radiation of the 
heater had sufficient intensity near the band-gap wavelengths 
to serve as a radiation source for the transmission 
measurement even at temperatures as low as 200$^\circ$C.

\begin{figure}
\includegraphics[width=8cm]{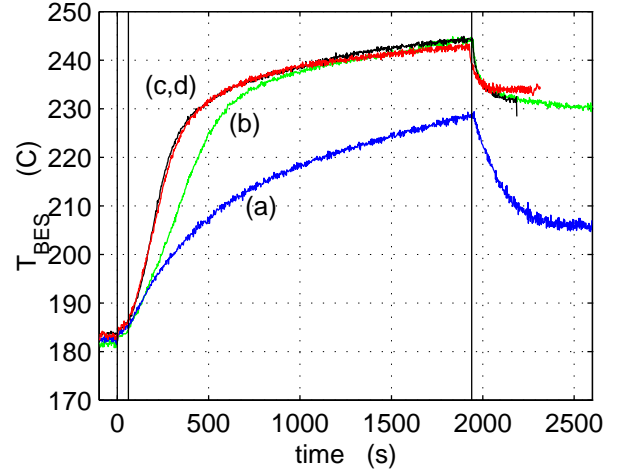}
\caption{
Temporal evolution of substrate temperatures measured 
by the band-edge spectroscopy. The grown layers differ by
the Mn contents of 
2.5\% (curve a), 3.5\% (curve b), 5\% and 7\% (curves c and d), 
respectively.
Times $t=0$ (opening the Ga source), $t=60$~s (opening the
Mn source), and $t=1950$~s (closing Ga and Mn sources) are 
indicated by vertical lines. 
}
\label{TBs}
\end{figure}

The low-temperature part of the structures under investigation 
consisted of 5~nm thick GaAs buffer and 150~nm 
thick $\rm{Ga_{1-x}Mn_xAs}$ layer, with x varying from 2.5\% to 7\%.
At the growth rate used the temperature of the Ga source
was 930$^\circ$C, the Mn source temperature varied from 770 to
830$^\circ$C. During the growth the substrate heater was 
supplied by a constant power, without any feedback temperature
control. The same initial substrate temperature 
of $182\pm 2^\circ$C was reproduced when
applying the same power and using always the same 
piece of sample holder.

In Figure~\ref{TBs} temporal evolutions of the BES-measured 
substrate temperatures ($T_{\rm BES}$) are shown for growths 
with various Mn densities. All the curves have 
several common features: 
{\it(i)} small (less than 5 degrees) increase of $T_{\rm BES}$ 
after opening the Ga source; 
{\it(ii)} pronounced (40--60 degrees) $T_{\rm BES}$ increase 
after opening the Mn source;
{\it(iii)} $T_{\rm BES}$ decrease after closing the sources,
with a new steady-state value well above the initial substrate
temperature.
During all the growths the temperature measured by usual floating 
thermocouple behind the substrate showed only a weak increase 
(less than 3 degrees) which relaxed back after the growths.

\begin{figure}
\includegraphics[width=8cm]{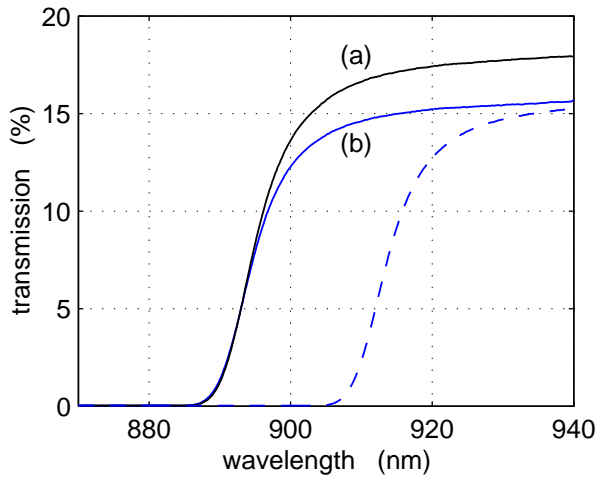}
\caption{
Near band gap transmission spectra of an undoped GaAs substrate 
(curve a), and the same substrate with a 50~nm thick GaMnAs
layer with 7\%~Mn (curve b). Both curves were measured at room
temperature. For contrast, curve (b) is repeated 
in the dashed line with a wavelength shift of 
$\Delta\lambda=19\,\mu$m which would correspond to increase 
of temperature from 
200 to 250~$^\circ$C (Ref.~\onlinecite{Shanabrook}).
}
\label{bgspectra}
\end{figure}

When considering the dramatic difference between the $T_{\rm BES}$
response to opening the Ga and Mn sources, a question arises
on the validity of the BES interpretation as a change in
substrate temperature. An alternative explanation could consist
in a change of the transmission spectrum shape due to
presence of Mn impurity levels close to the fundamental absorption 
edge. If it were so, a relevant difference would have to be 
detectable between spectra of an undoped GaAs and a GaMnAs
layer at the same temperature. Two such spectra are shown
in Fig.~\ref{bgspectra}, representing a semiinsulating GaAs 
substrate, and the same substrate with a GaMnAs layer 
on top of it. In the band edge region, however, there appears 
to be no significant difference between the spectra,
comparable to the expected shift due to difference in 
temperature.\cite{BESthermo,Urbach}
Thus, the chemically induced spectrum change is unlikely in the relevant 
parameter range and the observed band edge shift can indeed be 
attributed to the change in temperature. The more the question 
recurs on the mechanism of the dramatic temperature increase after 
opening the Mn source. 

\begin{figure}
\includegraphics[width=8cm]{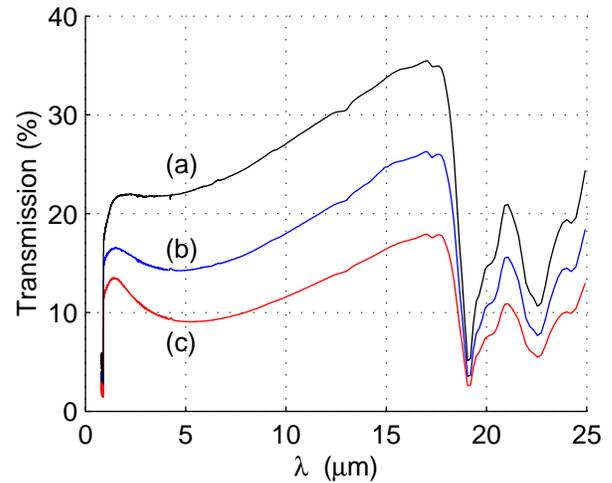}
\caption{
Infrared transmission spectra of undoped GaAs substrate
and GaMnAs layers grown on the same substrate. Parameters
of the samples are as follows: (a) undoped substrate;
(b) 50~nm, 7\%~Mn;
(c) 150~nm, 7\%~Mn.
All curves were measured at room temperature.
}
\label{irspectra}
\end{figure}

We attribute this effect to increasing heat absorption in 
the heavily doped epitaxial layer. Two straightforward absorption 
channels may be considered, related to the Mn doping:
excitation of valence electrons to the Mn acceptor level, and
free carrier absorption.
The first channel can readily be neglected: with the ionization energy 
around 100~meV (Ref. \onlinecite{Parker}) the Mn atoms are almost 
fully occupied at 
the room temperature, making further transitions from the valence 
band impossible. On the other hand, the absorption efficiency via 
free holes is strong due to their extremely high density (in the 
order of $10^{20}$cm$^{-3}$). Since, moreover, their mobility
is low and relaxation time short (in the order of 10~cm$^2$/Vs, and 
$10^{-16}$s, respectively), a significant absorption
can be expected in a broad infrared range, starting already in
the GaAs bandgap.\cite{Cardona} This is
illustrated in Fig.~\ref{irspectra} where transmission spectra
between 1 and 25~$\mu$m are shown for various samples.

Besides the free hole absorption the incoming heat is absorbed 
also by crystal lattice vibrations, independently of the doping.
A strong two-phonon absorption is seen in the long-wavelength 
part of Fig.~\ref{irspectra}. This absorption and its
one-phonon analog\cite{Blakemore} around 35~$\mu$m
are mainly responsible for heating of an insulating 
GaAs substrate.\cite{heating}

\section{Model}

It is tempting to describe the heat exchange in the substrate
simply as difference of {\it (i)} a heating term (proportional 
to heat flux from thermal sources, with an absorption factor 
increasing in time), and {\it (ii)} a cooling term (proportional
to deviation from a steady state temperature).
However, such a model turns out to be inadequate, yielding some
results contradictory to experiments.
For example, it predicts an unbounded increase of the substrate
temperature, as long as the absorption increases, i.e.~until 
the layer becomes completely opaque. Figures \ref{TBs} and 
\ref{irspectra} show a different behavior: 
the temperature stabilizes in spite of the nonsaturated
absorption. The reason of this failure is fundamental:
according to Kirchhoff's law the absorption efficiency of a body
changes identically to its emission efficiency at each 
wavelength,\cite{Lienhard} i.e. no temperature change
can occur unless the radiation spectrum changes.
Disregarding the spectral distribution in a system with significant 
radiative coupling, as in the above model, inevitably leads 
to nonphysical solutions.

A mathematical model we propose assumes fully radiative
heat exchange. Its essential ingredients are proper
description of the spectral power distributions 
of the thermal sources involved, and the doping induced 
change in the absorption and emission characteristics of the substrate.

A radiating and absorbing body can be approximated by
a black body with spectral radiance (monochromatic emissive power) 
$R(\lambda)$ enclosed by a surface with dimensionless emittance 
$\epsilon(\lambda)$ and absorptance $\alpha(\lambda)$.
The radiance $R(\lambda)$ obeys the Planck law,
\begin{equation}
R{(\lambda;T)} = \frac{2\pi hc^2}{\lambda^5}[\exp(\frac{hc}{\lambda k_BT})-1]^{-1}
\end{equation}

The net heat transferred from a radiating body 1 to an 
absorbing/radiating body 2 can then be described as
\begin{equation}
Q_{net} = F_{1-2}S_2
\int_0^\infty{[\alpha_2(\lambda)\epsilon_1(\lambda)R_1(\lambda)
               -\epsilon_2(\lambda)R_2(\lambda)]d\lambda}
\label{qnet12}
\end{equation}
where dimensionless factor $F_{1-2}$ 
represents the geometric configuration of the two bodies,
$S_2$ is the surface area of body 2.
This general principle can be used to describe --- pair
by pair --- the heat exchange between the substrate and 
all other relevant thermal sources in the growth chamber. 

To simplify the description we adopt the following assumptions:\\
({\it i})  
emittance of each body equals to its absorptance, 
$\epsilon(\lambda)=\alpha(\lambda)$ (diffuse form of Kirchhoff's 
law\cite{Lienhard});\\
({\it ii})  
the heater and the cells behave like gray bodies, i.e.~their 
emittances are constants;\\
({\it iii}) absorptance/emittance $\alpha_s(\lambda)$ of the substrate 
is composed of the phonon and the free carrier contributions.
For simplicity, the phonons are assumed to absorb perfectly in their 
wavelength interval(s), $(\lambda_{p1},\lambda_{p2})$,
the free carrier contribution $\alpha_{f}$ is assumed
to rise in time, starting from zero in the undoped substrate and
exponentially saturating as the total number of free holes increases, 
\begin{equation}
\alpha_s(\lambda) = \left\{
\begin{array}{ll}
1, & \lambda_{p1} < \lambda  < \lambda_{p2}\\
\alpha_{f}=1-\exp(-k_f t), & \lambda \quad {\rm elsewhere}
\end{array}
\right.
\label{alfa}
\end{equation}
where the rate constant $k_f$ is proportional to dopant flux.

The following bodies are relevant for the heat exchange:
the substrate heater (temperature $T_h$), the effusion cells
opened during the growth (effective temperature $T_c$), the background of
the growth chamber (effective temperature $T_b$), and the substrate
(temperature $T_s$). The view factor $F$ for each pair of bodies
can be approximated by the solid angles $A_h$, $A_c$,
$A_b$, at which the substrate with area $S_s$ "sees" the respective bodies;
identity $A_h=A_b+A_c$ holds, as the heater covers one complete hemisphere
of the substrate's scope ($A_h$), the other hemisphere being filled by the background ($A_b$)
with small hot spots of the opened cells ($A_c$).
The heat balance in the substrate is then
\begin{eqnarray}
Q_{net} = & S_s \sum\limits_i
\int_0^\infty{A_i[R(\lambda;T_i)-R(\lambda;T_s)]\,\alpha_s\,d\lambda},
\nonumber\\
& i=h, b, c
\label{qnetinteg}
\end{eqnarray}
Absorptance $\alpha_s(\lambda)$ is a piecewise constant function, with its
phonon part lying in a narrow interval in the mid infrared range. There the Planck's 
law approximates to $R \propto T/\lambda^4$ (Rayleigh-Jeans law), leading to 
linear terms in $T$ when integrating Eq.~(\ref{qnetinteg}), whereas 
the complete Planck's law yields terms proportional to $T^4$.
Thence, equation~(\ref{qnetinteg}) after integration results in
\begin{eqnarray}
Q_{net}& = & \sigma S_s \times \nonumber\\
&&\left\{\alpha_{f}\left[T_h^4+a_cT_c^4+(1-a_c)T_b^4-2T_s^4\right]+\right.\nonumber\\
&&\left.(1-\alpha_{f})\beta\left[T_h+a_cT_c+(1-a_c)T_b-2T_s\right]\right\}\nonumber\\
&&
\label{qnet}
\end{eqnarray}
where $\sigma$ is the Stefan-Boltzmann constant, $a_c=A_c/A_h$ is the view angle ratio 
of the opened cells, and $A_b/A_h=1-a_c$. The constant factor
$\beta$ reflects the ratio of power irradiated in the phonon
domain and power over all wavelengths,
\begin{equation}
\beta = \frac{5c^3h^3}{\pi^4k_B^3}\left(1/\lambda_{p2}^3-1/\lambda_{p1}^3\right)
\label{beta}
\end{equation}
Taking $\lambda_{p1}=16.5\,\mu$m and $\lambda_{p2}=18.5\,\mu$m 
leads to $\beta=1\times10^7\,\rm K^3$ as an order of magnitude 
estimate.\cite{myphonon}

In equilibrium, $Q_{net}=0$, and Eq.~(\ref{qnet}) determines the steady state
substrate temperature $T_s$ for any given $T_h$, $T_b$, $T_c$, and $\alpha_{f}$;
parameters $\beta$ and $a_c$ are physical and technical constants, respectively.
Out of equilibrium, the temporal evolution of $T_s$ is governed by
\begin{equation}
C dT_s/dt = Q_{net},
\label{dTdt}
\end{equation}
where $C$ is constant heat capacity of the GaAs substrate; varying 
electronic specific heat 
can safely be neglected at the relevant temperatures.\cite{Ashcroft} 
Let us note that if the change in temperature is slow enough
the system remains quasistationary and $T_s(t)$ can still be 
find alone by solving $Q_{net}=0$ with $\alpha_{f}(t)$ increasing
in time according to Eq.~(\ref{alfa}).

Explicit solutions can be obtained in two special stationary cases:\\
a) {\it Negligible free electron absorption}, or $\alpha_{f}=0$.
Equation~(\ref{qnet}) simplifies to
\begin{equation}
T_{s} = (T_h+a_cT_c+(1-a_c)T_b)/2
\label{Ts1}
\end{equation}
In particular, $T_s=(T_h+T_b)/2$ in the steady state before the growth, 
when $a_c=0$.\\
b) {\it Prevailing free electron absorption}, or 
$\alpha_{f}T_s^4 \gg (1-\alpha_{f})\beta T_s$. Then,
\begin{equation}
T_{s}^4 = \left(T_h^4+a_cT_c^4+(1-a_c)T_b^4\right)/2
\label{Ts4}
\end{equation}
In particular, $T_s^4=(T_h^4+T_b^4)/2$ in the steady state after 
the growth, explaining the persistent increase of $T_s$ as a consequence
of change of the absorption characteristic in presence of two different 
thermal sources.

\begin{figure}
\includegraphics[width=8cm]{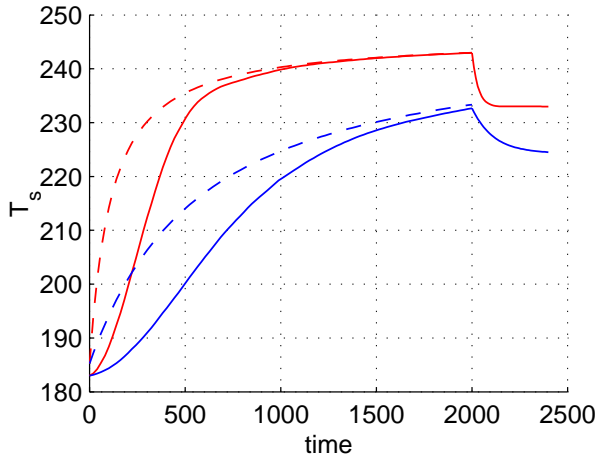}
\caption{
Calculated temperature evolutions of substrate temperatures
of strongly and weakly doped layers (solid lines). Also shown 
are corresponding quasistationary solutions (dashed lines).
}
\label{calcul}
\end{figure}

The above stationary solutions can be used to determine the model
parameters from the measured data. In accordance with Fig.~\ref{TBs} we
can take
$T_s=183^\circ$C before the growth, 
$T_s=243^\circ$C at the end of the growth of the most strongly doped layer, and
$T_s=233^\circ$C long after closing the cells. Taking further
$a_c=0.005$ as a rough estimate of the view angle ratio of 
the opened cells, we can solve Eqs.~(\ref{Ts1}) and (\ref{Ts4}),
getting $T_h=322^\circ$C for the substrate heater
temperature, $T_c=954^\circ$C for the effective temperature
of the opened cells, and $T_b=44^\circ$C for the effective
background temperature. This single set of
plausible model parameters yields quasistationary solutions
for $T_s$ shown in Fig.~\ref{calcul} by dashed lines, if
we chose $k_f=10\times 10^{-5}\rm s^{-1}$ ($2\times 10^{-5}\rm s^{-1}$)
as the doping rate constant of the strongly (weakly) doped layer. 
Trying, finally, $\sigma S_s C^{-1}=1\times 10^{-10}\rm K^{-3}s^{-1}$
(corresponding to 2-inch GaAs substrate of $500\,\mu$m thickness),
Eqs.~(\ref{qnet}) and (\ref{dTdt}) lead to time
evolutions $T_s(t)$ plotted by solid lines in Fig.~\ref{calcul}.

\section{Discussion}

On account of its simplifications the model cannot have
the ambition of a perfect agreement.
Its main virtue lies in the fact that it
links up parameters of a straightforward physical
meaning with the observed temperature evolution,
and that physically reasonable values of the model
parameters lead to expectable results. 

The accuracy of the agreement could be arbitrarily increased,
e.g., by including additional
thermal sources in the growth chamber and by tuning
their temperatures. This, however, would not necessarily
make the model more realistic. There are numerous influences
out of reach of a simple description. For example,
a part of the heat exchange is due to a conduction
between the sample holder and other massive parts of 
the holder mechanism, all of them with particular
heat capacities; the free electron absorption is wavelength
dependent in the IR range; all surfaces exposed to material
fluxes change their absorptance/emissivity during the growth,
leading to changes of their temperatures;
neither the phonon domain is absolutely absorbing,
nor the near infrared absorption is completely
absent in the undoped substrate; reflection takes
place, etc.

The increase in the 
substrate temperature is hardly visible for the floating
thermocouple behind the substrate, and a temperature
control based on it is thus inefficient. Yet, it is
desirable to keep the temperature change small,
especially in highly doped GaMnAs layers, where the growth
temperature is strongly constrained both from the 
bottom (disordered condensation) and the top (precipitation).
Several schemes can be applied to meet this goal:\\
1. BES-locked substrate heater control;\\
2. anticipatory throttling of the substrate heater;\\
3. In-bonding of the substrate to a massive sample holder
with inherently strong IR absorption;\\
4. incorporation of a less doped (uncritical in temperature)
but sufficiently absorbing GaMnAs sublayer between the GaAs 
substrate and the final GaMnAs structure.

\begin{figure}
\includegraphics[width=8cm]{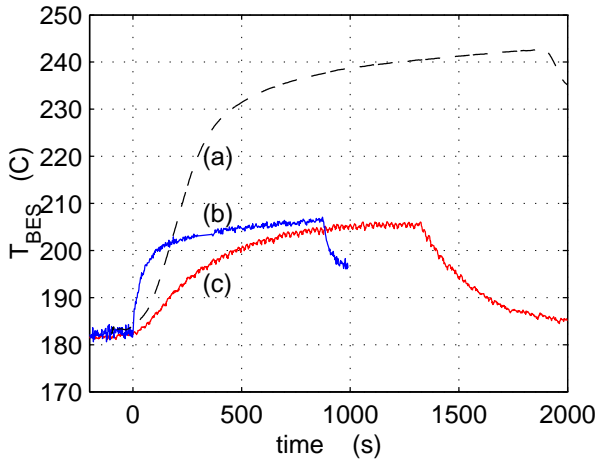}
\caption{
Substrate temperature evolution during growth of $x_{\rm Mn}=7\%$
layers on 
(a) the In-free mounted substrate,
(b) the In-free mounted sandwich structure with a buried 
GaMnAs heating layer,
(c) the In-bonded substrate.
The buried heating layer in (b) is 150~nm
thick with $x_{\rm Mn}=5\%$, separated by 25~nm GaAs from 
the top GaMnAs layer.
}
\label{TBs3}
\end{figure}

Temperature evolutions of the In-bonded substrate and 
the structure with the buried GaMnAs heating layer are 
shown in Fig.~\ref{TBs3}. The curves clearly differ in their
time constants, corresponding to different heat capacities.
The temperature increase, however, is almost identical in both 
systems, reduced to less than one half if compared
with Fig.~\ref{TBs}. 

This illustrates an important
difference between systems possessing a significant
IR absorptance (such as In-bonded substrates, or
sandwiched structures with buried heating layer), and
systems where a wide-range IR absorptance is missing
(such as In-free mounted semiinsulating substrates).
In the first case the temperature increase is solely due 
to activation of an additional thermal source; 
wide-range IR absorption is already prevailing and
its further change (if any) does not affect the
heat-exchange balance. In terms of the presented
model, terms linear in $T$ are negligible against
terms $T^4$.

In the second case the change of the {\it spectrum} of 
the substrate absorptance/emittance is essential;
the change in temperature corresponds to a shift of the
absorption maximum from longer to shorter wavelengths.
In terms of the model, at first the main heat exchange 
takes places through terms linear in $T$, later the
$T^4$ terms prevail. The corresponding temperature 
increase would occur even if no additional {\it heat} flux
appears. On the other hand, in a system with negligible
absorption towards the shorter wavelengths, opening a
high temperature cell with small orifice has little
or no effect on the absorbed heat. Thus, an interesting 
paradox can occur: temperature variations in an In-bonded 
substrate can turn {\it bigger} than those in an In-free
mounted substrate with weak free-carrier absorption.

\section{Summary}
 
The main part of the dramatic increase in the substrate 
temperature could unambiguously be connected with the
heavy Mn doping via the following scenario:\\
({\it i}) Before the growth, the undoped substrate
absorbs the radiation from the thermal sources mainly
in the mid-infrared range (phonon domain).\\
({\it ii}) The doping leads to strongly enhanced
free-carrier absorption in the near-infrared range,
shifting the decisive spectral absorptance
towards the shorter wavelengths.\\
({\it iii}) More heat is absorbed from the high-temperature
(short wavelength) sources. To reach a new heat-exchange 
equilibrium, the substrate radiance has to shift accordingly,
i.e.~its temperature increases.

The proposed model, in contrast to a linear cooling model, 
explains all the main features of
the observed effect, while preserving the involved
physical mechanisms disclosed. Using plausible values
of the input parameters it leads to quantitatively correct
results. It yields conclusions relevant also for 
temperature stability during the MBE growth generally.
Especially, it sheds a new light on behavior of In-bonded
and In-free mounted substrates.

\bigskip

\begin{acknowledgments}
We greatfully acknowledge stimulating discussions with 
R.P.~Campion, as well as his contribution to the early stages 
of this work.
The work was done in the framework of AV0Z1-010-914 program, 
supported by grants GACR 202/04/1519, ASCR KAN400100625,
MoE LC510, and FON/06/E002 (ESF Eurocores-FoNE 
ERAS-CT-2003-980409).
\end{acknowledgments}


\begin{thebibliography}
\bibitem{heating}
The interband absorption is negligible: a substrate heater at 300$^\circ$C
radiates by factor 10$^{-6}$ less power in the band gap region than
in the one and two phonon regions.
\bibitem{myphonon}
Used wavelengths correspond to two-phonon domain according to 
Ref.\onlinecite{Blakemore}; one phonon domain has been 
neglected. Using the somewhat longer wavelengths according to 
Fig.~\protect{\ref{irspectra}} one would obtain 
$\beta=6\times10^6\,\rm K^3$.
\end{thebibliography}
\end{document}